\documentclass[10pt]{amsart}

\usepackage{amsmath,amssymb,amsthm,a4,bm}

\newcommand{\Tm}{\sqrt{-\Delta + m^2} \, }
\newcommand{\RR}{\mathbb{R}}
\newcommand{\CC}{\mathbb{C}}
\newcommand{\NN}{\mathbb{N}}
\newcommand{\TPsi}{\bm{\Psi}}
\newcommand{\TPhi}{\bm{\Phi}}

\newcommand{\En}{\mathcal{E}}
\newcommand{\EnHF}{\mathcal{E}_{\mathrm{HF}}}
\newcommand{\EnH}{\mathcal{E}_{\mathrm{H}}}
\newcommand{\Nn}{\mathcal{N}}
\newcommand{\Enz}{\widetilde{\mathcal{E}}}
\newcommand{\dd}{\mathrm{d}}

\newcommand{\Hhalf}{H^{1/2}}
\newcommand{\Hs}{H^s}
\newcommand{\Hrhalf}{H_{\mathrm{rad}}^{1/2}}
\newcommand{\Ctest}{C^\infty_{c}}

\newcommand{\Ncr}{N_{\mathrm{cr}}}
\newcommand{\Mcr}{M_{\mathrm{cr}}}
\newcommand{\kapcr}{\kappa_{\mathrm{cr}}}

\newcommand{\ie}{i.\,e.}
\newcommand{\eg}{e.\,g.}

\newtheorem{theorem}{Theorem}
\newtheorem{lemma}{Lemma}

\newtheorem*{remark*}{Remark}
\newtheorem*{remarks*}{Remarks}

\begin{document}

\title[Dynamical Collapse of White Dwarfs]{
{\bf Dynamical Collapse of White Dwarfs in \\ Hartree- and Hartree-Fock Theory}
}

\author{J\"urg Fr\"ohlich and Enno Lenzmann}

\address{J\"urg Fr\"ohlich, Institute for Theoretical Physics, ETH Z\"urich--H\"onggerberg, 8093 Z\"urich, Switzerland}
\email{juerg@itp.phys.ethz.ch}

\address{Enno Lenzmann, Department of Mathematics, Massachusetts Institute of Technology, Room 2-230, Cambridge, MA 02139, U.S.A.} 
\email{lenzmann@math.mit.edu}

\date{August 7, 2006}

\maketitle


\begin{abstract}
We study finite-time blow-up for pseudo-relativistic Hartree- and Hartree-Fock equations, which are model equations for the dynamical evolution of white dwarfs. In particular, we prove that radially symmetric initial configurations with negative energy lead to finite-time blow-up of solutions. Furthermore, we derive a mass concentration estimate for radial blow-up solutions. Both results are mathematically rigorous and are in accordance with Chandrasekhar's physical theory of white dwarfs, stating that stellar configurations beyond a certain limiting mass lead to ``gravitational collapse'' of these objects. Apart from studying blow-up, we also prove local well-posedness of the initial-value problem for the Hartree- and Hartree-Fock equations underlying our analysis, as well as global-in-time existence of solutions with sufficiently small initial data, corresponding to white dwarfs whose stellar mass is below the Chandrasekhar limit.
\end{abstract}


\section{Introduction and Description of the Problem}

This paper is a contribution to the mathematical physics of white dwarfs and neutron stars. White dwarfs are dense stars composed of electrons and nuclei, which form a completely ionized plasma. The electrostatic Coulomb forces between these particles establish local electric neutrality to a high degree. For this reason, these forces are screened almost perfectly and make only a very modest contribution to the dynamical evolution and the energy of a white dwarf. Local electric neutrality implies that the spatial and momentum distributions of nuclei are approximately equal to those of the electrons. Since the masses of nuclei are much larger than the mass of an electron, the leading contribution to the total kinetic energy of a white dwarf comes from the electron gas, while the main contribution to its potential energy is due to the gravitational interaction among the nuclei. To simplify matters, we consider a single species of nuclei of electric charge $Ze$ and mass $m_Z \gg m$, where $-e$ is the electric charge of the electron and $m$ its mass. Throughout this paper, we use units such that Planck's constant $\hbar = 1$ and the velocity of light $c=1$.

\subsection{Heuristic Discussion}
Let $N$ denote the number of electrons in a white dwarf, and let $R$ be its radius. Then the number of nuclei is $N/Z$, and the average momentum, $p$, of a nucleus or electron is given by $p \simeq N^{1/3}/R$. A rough estimate for the groundstate energy, $E(N)$, of such a star is thus given by
\begin{equation} \label{eq:grE}
E(N) = \min_{p} \Big \{ N \sqrt{p^2 + m^2} + \frac{N}{Z} m_Z - \frac{1}{2} \left ( \frac{N}{Z} \right )^2 \frac{G m_Z^2}{R} \Big \} \Big |_{R=N^{1/3} p^{-1}} \, .
\end{equation}   
The constant $\frac{N}{Z} m_Z$ is the rest energy of the nuclei and will be subtracted from the groundstate energy henceforth; the constant $G$ is Newton's gravitational constant. The function of $p$ within curly brackets on the right side of (\ref{eq:grE}) has a minimum $ > -\infty$ only if
\begin{equation}
N < \Ncr := \big ( G m_Z \big )^{-3/2} Z^3,
\end{equation}
where $\Ncr$ is the {\em ``Chandrasekhar number.''} Thus, a white dwarf of total mass $M \simeq \frac{N}{Z} m_Z$ larger than the so-called {\em ``Chandrasekhar mass''}
\begin{equation}
 \Mcr := \frac{\Ncr}{Z} m_{Z},
\end{equation} 
is energetically unstable and is expected to undergo gravitational collapse; see \cite{Chandrasekhar1931}.

For $N \lesssim \Ncr$, the momentum $p$ minimizing the right side of (\ref{eq:grE}) is of the order of $m$; (more precisely, $p \gtrsim m (N/\Ncr)^{3/2}$, with $p \rightarrow \infty$, as $N \rightarrow \Ncr-$, see \eg~\cite{Straumann2004}). The use of relativistic kinematics for the electrons in a calculation of the groundstate energy of a white dwarf is therefore mandatory. Furthermore, for typical white dwarfs, the ratio between the Schwarzschild radius, $2GM$, and the radius $R$ of the star, as determined by (\ref{eq:grE}), is of the order of $10^{-4}$, so that effects of general relativity are unimportant, and gravity can be described by unretarded Newtonian two-body forces.

The outline of a heuristic description of neutron stars is similar, except that effects of general relativity become more important.

\subsection{Hartree- and Hartree-Fock Equations}
We now propose to describe a white dwarf, or a neutron star, quantum-mechnically, but within the approximation described above and assuming that the number $N$ of electrons is conserved. This leads us to consider the Hamilton operator
\begin{equation}
H^{(N)} = \sum_{k=1}^N \sqrt{p_k^2 + m^2} - \kappa \sum_{1 \leq k < l \leq N} \frac{1}{|x_k - x_l|}  \, ,
\end{equation}
where $p_k = -i \nabla_{x_k}$ and $\kappa = G m_Z^2/Z^2$, acting on the Hilbert space
\begin{equation}
\mathcal{H}^{(N)} = \big ( L^2(\RR^3) \otimes \CC^2 \big)^{\wedge N} .
\end{equation}
Here $L^2(\RR^3)$ is the space of square-integrable one-electron wave functions on physical space $\RR^3$, $\CC^2$ is the space of states of the spin of an electron, and {\em ``$\wedge \, N$''} denotes an $N$-fold antisymmetric tensor product, in accordance with the fact that electrons (and neutrons) are fermions, \ie, they satisfy the Pauli principle. In the following, electron spin plays a completely uninteresting r\^ole. To simplify our notation, we will therefore ignore it.

Special state vectors in the Hilbert space $\mathcal{H}^{(N)}$ are given by Slater determinants,
\begin{equation}
\triangle := \psi_1 \wedge \ldots \wedge \psi_N,
\end{equation}
where $\psi_1, \ldots, \psi_N$ are $N$ orthonormal one-particle wave functions; \ie,
\begin{equation} 
\langle \psi_k, \psi_l \rangle = \delta_{kl}, \quad \mbox{for all $k,l = 1, \ldots, N$,} 
\end{equation}
with $\langle \cdot, \cdot \rangle$ the scalar product on $L^2(\RR^3)$, (remember that we neglect electron spin). We note that the (dimensionless) coupling constant $\kappa = G m_Z^2 / Z^2$ is tiny, $\kappa \sim \mathcal{O}(10^{-38})$, while the number $N$ of electrons in a star is huge: $N \lesssim \mathcal{O}(10^{57})$. One expects that, in this regime, the groundstate of the Hamiltonian $H^{(N)}$ is well approximated by a Slater determinant, $\triangle_0$; see \cite{Lieb+Thirring1984, Lieb+Yau1987}. Moreover, the quantum-mechanical time evolution, as described by the one-parameter unitary group $\{ \exp(-it H^{(N)}) \}_{t \in \RR}$, is expected to evolve a Slater determinant, $\phi_1 \wedge \ldots \wedge \phi_N$, describing the state of the star at time $t=0$, to another Slater determinant,
\begin{equation} \label{eq:slater}
\psi_1(t) \wedge \ldots \wedge \psi_N(t),
\end{equation}
at time $t>0$, up to an error term that tends to $0$ in the {\em ``mean-field limit''} $\kappa \rightarrow 0$, with $N \sim \mathcal{O}(\kappa^{-3/2})$. In (\ref{eq:slater}), the one-particle wave functions $\psi_1(t), \ldots, \psi_N(t)$ are solutions of the $N$ coupled equations
\begin{equation} \label{eq:HF} \tag{HF}
i \partial_t \psi_k = \Tm \psi_k - \sum_{l=1}^N \big ( \frac{\kappa}{|x|} \ast |\psi_l|^2 \big ) \psi_k + \sum_{l=1}^N \psi_l \big ( \frac{\kappa}{|x|} \ast \{ \overline{\psi}_l \psi_k \} \big ) ,
\end{equation}
with initial conditions $\psi_k(t=0) = \phi_k$,  $k=1, \ldots, N$, and time $0 \leq t < T$, where $0 < T \leq \infty$ is the maximal time of existence of the solution $\{ \psi_k(t) \}_{k=1}^N$. In (HF), the pseudo-differential operator $\sqrt{-\Delta + m^2}$, which is defined by its symbol $\sqrt{p^2 + m^2}$ in Fourier space, describes the kinetic energy, including the rest energy, of an electron, and the symbol $\ast$ denotes convolution of functions on $\RR^3$. The last term on the right side of (HF) is the so-called {\em ``exchange term,''} which is a consequence of the Pauli principle. When compared to the second term on the right side of (HF), the {\em ``direct term,''} it is subleading in the mean-field limit ($N \rightarrow \infty$). It is therefore often neglected. Then equation (HF) is replaced by the $N$ coupled equations
\begin{equation} \label{eq:H} \tag{H}
i \partial_t \psi_k = \Tm \psi_k - \sum_{l=1}^N \big ( \frac{\kappa}{|x|} \ast |\psi_l|^2 \big ) \psi_k ,
\end{equation}
with $\psi_k(t=0)= \phi_k$ and $k=1, \ldots, N$.

Equations (HF) are called (dynamical) {\em Hartree-Fock equations}, while (H) are called {\em Hartree equations.} These systems of evolution equations are the main characters studied in this paper.

The logics leading from the quantum-mechanical time evolution generated by the Hamiltonian $H^{(N)}$ to the nonlinear evolution equations (HF) and (H) for $N$ orthonormal one-particle wave functions, $\psi_1(t), \ldots, \psi_N(t)$, in the mean-field limit, $\kappa \rightarrow 0$ with $N = \mathcal{O}(\kappa^{-3/2})$, has been studied in \cite{EES2004}.

\subsection{Hamiltonian Structure and Conserved Quantities}
Let $\Hhalf(\RR^3)$ denote the inhomogeneous Sobolev space of index $1/2$. We define
\begin{equation}
\Gamma^{(N)} = \big (\Hhalf(\RR^3) \big )^{\times N} .
\end{equation}
This space can be interpreted as an affine {\em ``classical phase space''} with complex coordinates $\TPsi = (\psi_1, \ldots, \psi_N)$ and $\overline{\TPsi} = (\overline{\psi}_1, \ldots, \overline{\psi}_N)$. The symplectic 2-form, $\omega$, is given by
\begin{equation}
\omega = \frac{i}{2} \sum_{k=1}^N \dd \psi_k \wedge \dd \overline{\psi}_k .
\end{equation} 
We define a Hamilton functional, $\mathcal{H}_\#^{(N)}$, on $\Gamma^{(N)}$ by setting
\begin{equation}
\mathcal{H}_\#^{(N)}(\TPsi, \overline{\TPsi}) = \En_\#(\TPsi),
\end{equation}
with \# = HF or H, respectively, where the energy functionals $\EnHF$ and $\EnH$ are defined in Section \ref{sec:main}, below. Then eqs.~(HF) and (H) turn out to be the Hamiltonian equations of motion corresponding to the Hamilton functionals $\mathcal{H}_{\mathrm{HF}}^{(N)}$ and $\mathcal{H}_\mathrm{H}^{(N)}$, respectively. Formally, the quantities 
\[
\En_\#(\TPsi(t)) \quad \mbox{and} \quad \langle \psi_k(t), \psi_l(t) \rangle,
\]
with $k,l=1, \ldots, N$, are conserved under the Hamiltonian flow determined by eqs.~(HF), (H), respectively. These conservation laws play an important r\^ole in our analysis.

\subsection{Notation}
Throughout this text, we make use of inhomogeneous and homogenous Sobolev spaces of order $s$, denoted by $H^s(\RR^3)$ and $\dot{H}^s(\RR^3)$, which are equipped with norms $\| u \|_{H^s} = \| (1+\sqrt{-\Delta})^{s/2} u \|_{L^2}$ and $\| u \|_{\dot{H}^s} = \| (-\Delta)^{s/2} u \|_{L^2}$, respectively. The scalar product on $L^2(\RR^3)$ is defined as $\langle u,v \rangle = \int_{\RR^3} \overline{u} \, v \, \dd x$. 

With some abuse of notation, we sometimes identify collections of wave functions, $\TPsi = \{ \psi_k \}_{k=1}^N$, with ordered tuples $\vec{\TPsi}=(\psi_1, \ldots, \psi_N)$. For solutions of (HF) and (H), this procedure is legitimate, thanks to their $U(N)$-gauge invariance; see Section \ref{sec:main}, below.
 
In what follows, we write $X \lesssim Y$ if $X \leq C Y$, where $C$ is some universal constant. With regard to physical applications of (H) and (HF), we remind the reader that we use units such that $\hbar = c = 1$.


\section{Main Results} \label{sec:main}

We begin by reviewing some aspects of equations (H) and (HF). First, we recall from Section 1 that both sets of coupled equations exhibit (formally, at least) conservation of energy. That is, the {\em Hartree energy,}    
\begin{equation}
\EnH(\TPsi) = \sum_{k=1}^N \langle \psi_k, \Tm \psi_k \rangle - \frac{\kappa}{2} \int_{\RR^3} \! \int_{\RR^3} \frac{\rho_{\TPsi}(x) \rho_{\TPsi}(y)}{|x-y|} \, \dd x \, \dd y,
\end{equation}
and the {\em Hartree-Fock energy,}
\begin{equation}
\EnHF(\TPsi) = \sum_{k=1}^N \langle \psi_k, \Tm \psi_k \rangle - \frac{\kappa}{2} \int_{\RR^3} \! \int_{\RR^3} \frac{\rho_{\TPsi}(x) \rho_{\TPsi}(y) - |\rho_{\TPsi}(x,y)|^2}{|x-y|} \, \dd x \, \dd y,
\end{equation}
are conserved for solutions $\TPsi = \{ \psi_k \}_{k=1}^N$ of (H) and (HF), respectively. Here and in what follows, we make use of the {\em density matrix,}
\begin{equation}
\rho_{\TPsi}(x,y) = \sum_{k=1}^N \psi_k(x) \overline{\psi}_k(y), 
\end{equation}
and the {\em particle density,}
\begin{equation}
\rho_{\TPsi}(x) = \rho_{\TPsi}(x,x). 
\end{equation}
In addition to conservation of energy, we also have conservation of the {\em particle number} (proportional to the stellar mass) given by
\begin{equation}
\Nn(\TPsi) =  \int_{\RR^3} \rho_{\TPsi}(x) \, \dd x.
\end{equation} 
As we will see below, the conservation of $\Nn(\TPsi)$ is a special consequence of the $U(N)$-gauge symmetry; \ie, every transformation $\psi_k \mapsto \sum_{l=1}^N T_{kl} \psi_l$, with $T \in U(N)$, yields another solution of (H) and (HF), respectively.

\subsection{Initial-Value Problem}

Our first main result states local well-posedness of the initial-value problems for (H) and (HF), provided that the set of initial data belongs to $H^s(\RR^3)$, for some $s \geq 1/2$.  

\begin{theorem}[{\bf Local Well-Posedness}] \label{th:lwp}
Let (\#) denote either (H) or (HF). Suppose that $s \geq 1/2$ and let $N \geq 1$ be an integer. Then the initial-value problem for (\#) is locally well-posed in $\Hs(\RR^3)$.

By this we mean the following. For every collection of initial data, $\TPhi = \{ \phi_k \}_{k=1}^N \subset \Hs(\RR^3)$, there exists a unique solution, $\TPsi(t) = \{ \psi_k(t) \}_{k=1}^N \subset \Hs(\RR^3)$, solving (\#) such that
\[
\psi_k(0) = \phi_k \quad \mbox{and} \quad \psi_k \in C^0 \big ( [0,T); \Hs(\RR^3) \big ) \cap C^{1} \big ( [0,T); H^{s-1}(\RR^3) \big ) 
\]
holds, for all $k=1, \ldots, N$. Here $0 < T \leq \infty$ denotes the maximal time of existence, and $T < \infty$ implies that $\lim_{t \rightarrow T-} \| \psi_k(t) \|_{H^{1/2}} = \infty$ holds, for some $k=1, \ldots, N$.

In addition, the solution $\TPsi(t)$ depends continuously on $\TPhi$, and 
\[
\En_{\#}(\TPsi(t)) = \En_{\#}(\TPhi) \quad \mbox{and} \quad \Nn(\TPsi(t)) = \Nn(\TPhi)
\]
hold for all times $0 \leq t < T$. Moreover, we have that 
\[ \langle \psi_k(t), \psi_l(t) \rangle = \langle \phi_k, \phi_l \rangle , \]
for all $1 \leq k,l \leq N$ and all times $0 \leq t < T$.
\end{theorem}

\begin{remark*}{\em
The proof of this theorem proceeds along the lines of \cite{Lenzmann2006} and will only be sketched in Subsection \ref{sec:th:lwp}, below. We remark that no use of Strichartz-type estimates for the propagator, $e^{-it \sqrt{-\Delta + m^2}}$, is made throughout this proof. Note that when considering initial data below the energy norm, \eg, belonging to $H^s(\RR^3)$, for some $s < 1/2$, one would have to resort to such estimates; see \cite{Cho+Ozawa2006}. Since we are only interested in finite-energy solutions of (H) and (HF), we have no reason to pursue this issue here. }
\end{remark*}

The next theorem shows that sufficiently small initial data lead to global-in-time solutions. The smallness condition corresponds to a number of particles below the Chandrasekhar number, $\Ncr$, mentioned in Section 1. 

\begin{theorem}[{\bf Global solutions for $N \lesssim \Ncr$}] \label{th:gwp}
Every solution of either (H) or (HF), given by Theorem \ref{th:lwp}, exists for all times, $0 \leq t < \infty$, whenever the corresponding initial data, $\TPhi = \{ \phi_k \}_{k=1}^N \subset \Hs(\RR^3)$, form a collection of $L^2$-orthonormal functions whose number $N = \Nn(\TPhi)$ satisfies 
\[
N < \left ( \frac{\kapcr}{\kappa} \right )^{3/2} \, .
\]  
Here $\kapcr > 0$ is a universal constant of order 1.
\end{theorem}

\begin{remark*} {\em
If, in addition, the initial data $\TPhi$ satisfy $\En_\#(\TPhi) < Nm$ (the rest energy of the electrons) then $\| \TPsi(t) \|_{L^p}$ does not tend to 0, as $t$ tends to $\infty$, for any $p > 2$. The physical interpretation of this result is that $\TPsi(t)$ describes the evolution of a bound configuration of matter forming a star-like object. For details and a proof of a closely analogous result, see \cite{LenzmannDiss2006}. }
\end{remark*}

\subsection{Finite-Time Blow-Up}

We now turn our attention to {\em finite-time blow-up} for the Hartree equation (H), which, by Theorem \ref{th:gwp}, is only encountered for sufficiently large initial data. 

\begin{theorem}[{\bf Radial blow-up for (H) with negative energy}] \label{th:blowup}
Let $\TPhi = \{ \phi_k \}_{k=1}^N \subset \Ctest(\RR^3)$ be a collection of functions with the property that $\rho_{\TPhi}(x) = \sum_{k=1}^N |\phi_k(x)|^2$ is radially symmetric. If the Hartree energy is strictly negative, \ie,
\[ \EnH(\TPhi) < 0 , \] 
then the solution, $\TPsi(t) = \{ \psi_k(t) \}_{k=1}^N$, of (\ref{eq:H}) with initial data $\TPhi$ blows up within finite time. That is, we have that 
\[ \lim_{t \rightarrow T-} \| \psi_k(t) \|_{H^{1/2}} = \infty, \]
for some $k=1, \ldots, N$ and some $T < \infty$.  
\end{theorem}

\begin{remarks*} {\em
1) It is not difficult to construct an initial configuration $\TPhi$ with $\EnH(\TPhi) < 0$, as follows: We consider a ball in $\RR^3$ of radius $R > 0$ centered at the origin. We then pick an integer $N > 0$ and let $\phi_1, \ldots, \phi_N$ denote eigenfunctions of the Laplacian with Dirichlet boundary conditions at the boundary of the ball corresponding to the lowest $N$ eigenvalues and spanning a rotation-invariant subspace of $L^2(\RR^3)$. In accordance with the Pauli Principle, we choose $\TPhi := \{ \phi_k \}_{k=1}^N$. The relativistic kinetic energy of this configuration is proportional to $N^{4/3}$, while the gravitational potential energy is proportional to $-N^{2}$. Thus, if $N$ is sufficiently large then $\EnH(\TPhi)$ is strictly negative.  

2) We have also found an analogous blow-up result for (HF), but with the additional assumption that {\em each} function $\phi_k(x)$ has to be spherically symmetric. From the physical point of view such a hypothesis appears unnaturally strong, so that we refrain from formulating this blow-up result for (HF) as a theorem.    

3) The requirement that $\phi_k \in \Ctest(\RR^3)$ can be relaxed to weaker conditions on regularity and spatial decay. For the sake of simplicity of our presentation, we will not pursue this issue here. 

4) By invariance of $\TPhi = \{ \phi_k \}_{k=1}^N$ under spatial rotations we mean that, for every $R \in SO(3)$, $\phi_k(Rx) = \sum_{l=1}^N T_{kl}(R) \phi_l(x)$ holds, where $T(R) \in U(N)$ is some unitary matrix. This  implies in particular that the density matrix obeys $\rho_{\TPhi}(x,y) = \rho_{\TPhi}(Rx,Ry)$, for any $R \in SO(3)$. Moreover, we remark that it is easy to see that the corresponding unique solution, $\TPsi(t)$, of (HF) or (H) is also invariant under spatial rotations, for all times, $0 \leq t < T$, provided that $\TPsi(0) = \TPhi$ has this property.   
}
\end{remarks*}

Our last result shows that, when approaching the time of blow-up, any radial blow-up solution of (H) or (HF) exhibits a concentration of particles at the origin, whose number is at least of order of the Chandrasekhar number $\Ncr$.

\begin{theorem}[{\bf Chandrasekhar mass concentration for radial blow-up}] \label{th:Chandra}  
Let $\TPsi(t) = \{ \psi_k(t) \}_{k=1}^N$ be an $H^{1/2}$-valued solution of either (\ref{eq:H}) or (\ref{eq:HF}) that blows up at time $T >0$. Moreover, suppose that $\rho_{\TPsi(0)}(x) = \sum_{k=1}^N |\psi_k(0,x)|^2$ is radially symmetric. Then, for every $R > 0$, we have that
\begin{equation*}
\liminf_{t \rightarrow T-} \int_{|x| < R} \rho_{\TPsi(t)}(x) \, \dd x \geq \left ( \frac{\kapcr}{\kappa} \right )^{3/2} \, ,
\end{equation*}
where $\rho_{\TPsi(t)}(x) = \sum_{k=1}^N |\psi_k(t,x)|^2$. Here $\kapcr > 0$ is the same universal constant as in Theorem \ref{th:gwp}.
\end{theorem}

\begin{remark*}{\em
In view of the physical interpretation of (HF) and (H) discussed in Section 1, it would be of considerable interest to gain more insight into the properties of blow-up solutions for these equations and to arrive at a state of affairs comparable to what is known about blow-up for nonlinear Schr\"odinger equations (NLS) with $L^2$-critical, focusing nonlinearities; see, \eg, the monograph \cite{Cazenave2003} and references given there; (see, in particular, \cite{Merle+Tsutsumi1990, Weinstein1989} for mass concentration of blow-up solutions for NLS).}
\end{remark*}


\section{Proof of Main Results}

The proofs of Theorems \ref{th:lwp}--\ref{th:blowup} are extensions of arguments derived in \cite{Lenzmann2006, Froehlich+Lenzmann2006}, showing local and global well-posedness, as well as finite-time blow-up for the pseudo-relativistic Hartree equation, \ie, the equation
\begin{equation}
i \partial_t \psi = \sqrt{-\Delta + m^2} \, \psi - \big ( |x|^{-1} \ast |\psi|^2 ) \psi,
\end{equation}
where $\psi : [0,T) \times \RR^3 \rightarrow \CC$. We therefore only sketch the proofs of Theorems \ref{th:lwp}--\ref{th:blowup} in Subsections \ref{sec:th:lwp}--\ref{sec:th:blowup}. In contrast, the proof of Theorem \ref{th:Chandra} is given in detail in Subsection \ref{sec:th:Chandra}.

\subsection{Proof of Theorem \ref{th:lwp}} \label{sec:th:lwp}

For definiteness we consider the initial-value problem for (HF) in $\Hs(\RR^3)$, and we observe that all arguments apply to (H), with almost no change.

The initial-value problem for (HF) can be written as follows.
\begin{equation} \label{eq:ivp} \tag{IVP}
\left \{ \begin{array}{l} i \partial_t \vec{\TPsi} = \Tm \vec{\TPsi} + \kappa{\vec{\bm{F}}(\vec{\TPsi}}) , \\
\vec{\TPsi}(0) = \vec{\TPhi} \in H^{s,N}, \quad 0 \leq t < T. 
\end{array} \right .
\end{equation}
Here 
\begin{equation}
H^{s,N}:= \big ( \Hs(\RR^3) \big )^{\times N}
\end{equation}
is the $N$-fold cartesian product of $\Hs(\RR^3)$, equipped with the norm 
\begin{equation}
\| \vec{\TPhi} \|_{H^{s,N}} = \big ( \sum_{k=1}^N \| \phi_k \|_{H^{s}}^2 \big )^{1/2}.
\end{equation}
With some abuse of notation, we sometimes identify $\TPsi = \{ \psi_k \}_{k=1}^N$ with the vector $\vec{\TPsi}$, and, likewise, $\TPhi$ with $\vec{\TPhi}$. In (\ref{eq:ivp}) the nonlinearity, $\vec{\bm{F}} = (F_1, \ldots, F_N)$, is given by 
\begin{equation} \label{eq:Fk}
(F_k(\vec{\TPsi})) = - \sum_{l=1}^N \big ( |x|^{-1} \ast |\psi_l|^2 \big ) \psi_k + \sum_{l=1}^N \psi_l \big ( |x|^{-1} \ast (\overline{\psi}_l \psi_k) \big ).
\end{equation}

\begin{lemma} \label{lem:loclip}
Suppose $s \geq 1/2$, and let $N \geq 1$ be an integer. Then $\vec{\bm{F}} : H^{s,N} \rightarrow H^{s,N}$ is locally Lipschitz such that
\begin{equation} \label{eq:lem1}
\| \vec{\bm{F}}(\vec{\TPsi}) - \vec{\bm{F}}(\vec{\TPhi}) \|_{H^{s,N}} \lesssim \big ( \| \vec{\TPsi} \|_{H^{s,N}}^2 + \| \vec{\TPhi} \|_{H^{s,N}}^2 \big ) \| \vec{\TPsi} - \vec{\TPhi} \|_{H^{s,N}} ,
\end{equation}
\begin{equation} \label{eq:lem2}
\| \vec{\bm{F}}(\vec{\TPsi}) \|_{H^{s,N}} \lesssim \| \vec{\TPsi} \|^2_{H^{r,N}} \| \vec{\TPsi} \|_{H^{s,N}},
\end{equation}
for all $\vec{\TPsi}, \vec{\TPhi} \in H^{s,N}$, where $r = \max \{s-1,1/2 \}$.
\end{lemma}

\begin{remark*} {\em
The proof of Lemma \ref{lem:loclip} for the first term in (\ref{eq:Fk}) follows from a straightforward extension of \cite[Lemma3]{Lenzmann2006}, where an analogous result is shown for the nonlinearity $J : \Hs(\RR^3) \rightarrow \Hs(\RR^3)$ with $J(u) = (|x|^{-1} \ast |u|^2)u$, corresponding to Hartree nonlinearities and $N=1$. Also, the proof of the estimates (\ref{eq:lem1}), (\ref{eq:lem2}) for the second term in (\ref{eq:Fk}), \ie, the ``exchange term'', can be shown in a similar fashion.    }
\end{remark*}

By Lemma \ref{lem:loclip}, local-in-time existence and uniqueness of $\vec{\TPsi}(t) \in H^{s,N}$, as well as continuous dependence on $\vec{\TPhi}$, now follow by standard methods for evolution equations with locally Lipschitz nonlinearities; see \cite{Lenzmann2006} and references given there. In addition, estimate (\ref{eq:lem2}) and Gronwall's inequality allow us to deduce that, for any $s > 1/2$, 
\begin{equation*}
\sup_{0 \leq t \leq T_*} \| \vec{\TPsi}(t) \|_{H^{s,N}} \leq C ( T_*, \| \vec{\TPsi}(0) \|_{H^{s,N}}, \sup_{0 \leq t \leq T_*} \| \vec{\TPsi}(t) \|_{H^{1/2,N}} ), \quad \mbox{for $0 \leq T_* < T$}.
\end{equation*}
In particular, this implies that the maximal time of existence of any $\Hs$-valued solution of (HF), with $s > 1/2$, coincides with its maximal time of existence when viewed as an $H^{1/2}$-valued solutions; see also \cite{Lenzmann2006}.

Finally, we note that $U(N)$-charge conservation is due to $\frac{\dd}{\dd t} \langle \psi_k(t), \psi_l(t) \rangle =  0$, which follows from a direct calculation. Moreover, conservation of energy stems from the fact that $\frac{\dd}{\dd t} \En(\vec{\TPsi}(t)) = 0$ holds whenever the initial data satisfy $\vec{\TPhi} \subset H^{1,N}$. To prove conservation of energy for $H^s(\RR^3)$-valued solutions when $1/2 \leq s < 1$, one can proceed in a standard way, \ie, by using the continuous dependence on  initial data in $H^{s,N}$ and by appealing to the density of $H^{1,N} \subset H^{s,N}$. This completes our sketch of the proof of Theorem \ref{th:lwp}. \hfill $\blacksquare$

\subsection{Proof of Theorem \ref{th:gwp}} \label{sec:th:gwp}

Let $(\#)$ either stand for (H) or (HF). Then, by our hypothesis on the initial data and Theorem \ref{th:lwp}, we have that the solution, $\TPsi(t) = \{ \psi_k(t) \}_{k=1}^N \subset H^{1/2}(\RR^3)$, of (\#) forms a collection of $L^2$-orthonormal functions. By Lemma 2 (in Appendix A) and the estimate
\begin{equation}
\int_{\RR^3} \! \int_{\RR^3} \frac{ \rho(x) \rho(y) }{|x-y|} \, \dd x \, \dd y \leq C \big ( \int_{\RR^3} \rho(x) \, \dd x \big )^{2/3} \big ( \int_{\RR^3} \rho(x)^{4/3} \, \dd x ),
\end{equation}
for some constant $C \simeq 1$ (see \cite{Lieb+Yau1987} and reference given there), we obtain that
\begin{align}
\En_\#(\Psi) & \geq \sum_{k=1}^N \| \psi_k(t) \|_{\dot{H}^{1/2}}^2 - \frac{\kappa}{2} \int_{\RR^3} \! \int_{\RR^3} \frac{\rho_{\TPsi(t)}(x) \rho_{\TPsi(t)}(y)}{|x-y|} \, \dd x \, \dd y \nonumber \\
& \geq \Big [ 1 - \frac{\kappa}{\kapcr} \big ( \int_{\RR^3} \rho_{\TPsi(t)} (x) \, \dd x  \big )^{2/3} \big ] \sum_{k=1}^N \| \psi_k(t) \|_{\dot{H}^{1/2}}^2 ,
\end{align}
for some universal constant $\kapcr$ of order 1. Using that $\En_\#(\TPsi(t)) = \En_\#(\TPhi)$ and $\int_{\RR^3} \rho_{\TPsi(t)} = N$, we deduce the a-priori bound
\begin{equation}
\sup_{0 \leq t < T} \sum_{k=1}^N \| \psi_k(t) \|_{H^{1/2}}^2 \lesssim N + \big ( 1- \frac{\kappa}{\kapcr} N^{2/3} \big )^{-1} \En_\#(\TPhi),
\end{equation}
provided that $N < (\kapcr/\kappa)^{3/2}$ holds. By Theorem \ref{th:lwp}, this implies that the maximal time of existence is $T=\infty$. \hfill $\blacksquare$

\subsection{Proof of Theorem \ref{th:blowup}} \label{sec:th:blowup}

The proof of Theorem \ref{th:blowup} follows \cite{Froehlich+Lenzmann2006}, with some cosmetic changes only. We remark that smoothness and sufficient spatial decay of the initial data (e.\,g., that $\phi_k \in H^2(\RR^3)$ and $|\langle \phi_k, |x|^4 \phi_k \rangle| < \infty$) guarantee that all quantities involved in the following calculations are well defined.
 
First, we notice that 
\begin{equation}
a(t) := \sum_{k=1}^N \langle \psi_k(t), A \psi_k(t) \rangle, \quad \mbox{with $A:= -\frac{i}{2} ( x \cdot \nabla + \nabla \cdot x)$,}
\end{equation}
is found to satisfy the differential inequality
\begin{equation} \label{eq:blowup1}
\dot{a}(t) \leq \EnH(\TPsi(t)) = \EnH (\TPhi) , \quad \mbox{for $0 \leq t < T$.}
\end{equation}
Further, a calculation similar to the one in \cite{Froehlich+Lenzmann2006} shows that
\begin{equation}
m(t) := \sum_{k=1}^N \langle \psi_k(t), M \psi_k(t) \rangle, \quad \mbox{with $M := x \cdot \sqrt{-\Delta + m^2} x$},
\end{equation}
obeys
\begin{equation} \label{eq:blowup2}
\dot{m}(t) \leq 2 a(t) + C_1, \quad \mbox{for $0 \leq t < T$,}
\end{equation}
where $C_1 > 0$ is some constant only depending on $\Nn(\TPhi)$. In our proof of (\ref{eq:blowup2}), we make use of Newton's theorem for radially symmetric densities $\rho_{\TPsi(t)}(x)$, which forces us to assume that $\rho_{\TPhi}(x)$ be radially symmetric; see also Remark 4) following Theorem \ref{th:blowup}.

By combining (\ref{eq:blowup1}) and (\ref{eq:blowup2}) and integrating, we thus obtain 
\begin{equation} \label{eq:blowup3}
m(t) \leq \EnH(\TPhi) t^2 + C_1 t + C_2, \quad \mbox{for $0 \leq t < T$.}
\end{equation}
Since $m(t)$ is a nonnegative quantity, we conclude that if $\EnH(\TPsi(t)) = \EnH(\TPhi) < 0$ then the maximal time of existence $T$ must be finite. By Theorem \ref{th:lwp}, the solution, $\TPsi(t)$, of (H) thus blows up at time $T$, which is bounded from above by the positive root of the right-hand side of (\ref{eq:blowup3}). \hfill $\blacksquare$

\subsection{Proof of Theorem \ref{th:Chandra}} \label{sec:th:Chandra}
We consider (HF); but the proof for (H) is almost identical to the one presented below.
 
The proof makes use of variational arguments; see also \cite{Merle+Tsutsumi1990,Weinstein1989} for a proof of mass concentration for NLS. We argue by contradiction, as follows. Suppose the conclusion of Theorem \ref{th:Chandra} were not true. That is, there exists some $R > 0$ such that
\begin{equation} \label{eq:assChandra}
\lim_{n \rightarrow \infty} \int_{|x| < R} \rho_{\TPsi(t_n)}(x) \, \dd x < \left ( \frac{\kapcr}{\kappa} \right )^{3/2}
\end{equation} 
holds, where $\{ t_n \}_{n \in \NN}$ is a sequence of times increasing to $T$, as $n \rightarrow \infty$. Here $\kapcr > 0$ is the same universal constant as in Theorem \ref{th:gwp}.

To prove that assumption (\ref{eq:assChandra}) cannot hold, we begin by introducing a collection of $N$ sequences, $\bm{\widetilde{\Psi}}_n = \{ \widetilde{\psi}_{k,n} \}_{k=1}^N$, where
\begin{equation} \label{eq:Chandrapsi}
\widetilde{\psi}_{k,n}(x) := \sigma(t_n)^{-3/2} \psi_k(t_n, \sigma(t_n)^{-1} x ) .
\end{equation}
Here the strictly positive function $\sigma(t)$, defined on $[0,T)$, is given by 
\begin{equation}
\sigma(t) := \sum_{k=1}^N \| \psi_k(t) \|_{\dot{H}^{1/2}}^2 .
\end{equation}
Note that, since $\|\psi_k(t) \|_{L^2} = \| \psi_k(0) \|_{L^2}$, we have that $\sigma(t) \rightarrow \infty$, as $t \rightarrow T-$. Next, we define the functional
\begin{equation} \label{eq:Enz}
\Enz(\bm{\Phi}) := \sum_{k=1}^N \| \phi_k \|_{\dot{H}^{1/2}}^2 - \frac{\kappa}{2} \int_{\RR^3} \! \int_{\RR^3} \frac{\rho_{\bm{\Phi}}(x) \rho_{\bm{\Phi}}(y) - | \rho_{\bm{\Phi}}(x,y)|^2}{|x-y|} \, \dd x \, \dd y, 
\end{equation}
for any collection $\bm{\Phi} = \{ \phi_k \}_{k=1}^N \subset \Hhalf(\RR^3)$. A simple calculation then yields
\begin{equation} 
\Enz( \bm{\widetilde{\Psi}}_n) = \sigma(t_n)^{-1} \Enz(\bm{\Psi}(t_n)) .
\end{equation}
Moreover, by using conservation of mass and energy, it follows that
\begin{equation} \label{eq:Enz2}
| \Enz(\TPsi(t_n)) | \leq  |\En( \TPsi(0) ) | +  m \Nn(\TPsi(0)) 
\end{equation}
where we use that $0 \leq \langle \phi, \sqrt{-\Delta + m^2} \phi \rangle - \| \phi \|_{\dot{H}^{1/2}}^2 \leq m \| \phi \|_{L^2}^2$. Combining estimate (\ref{eq:Enz2}) with the fact that $\sigma(t_n)^{-1} \rightarrow 0$, as $n \rightarrow \infty$, we conclude that
\begin{equation} \label{eq:Enzzero}
\lim_{n \rightarrow \infty} \Enz(\bm{\widetilde{\Psi}}_n) = 0.
\end{equation}
For later use, we record that this implies that
\begin{equation} \label{eq:hls}
\lim_{n \rightarrow \infty} \int_{\RR^3} \! \int_{\RR^3} \frac{ \rho_{\bm{\widetilde{\Psi}}_n}(x) \rho_{\bm{\widetilde{\Psi}}_n}(y) - |\rho_{\bm{\widetilde{\Psi}}_n}(x,y)|^2}{|x-y|} \, \dd x \, \dd y= \frac{2}{\kappa} \, ,
\end{equation}
since $\sum_{k=1}^N \| \widetilde{\psi}_{k,n} \|_{\dot{H}^{1/2}}^2 = 1$ for all $n \in \mathbb{N}$, by construction of $\widetilde{\psi}_{k,n}$.

As a next step, we notice that $\bm{\widetilde{\Psi}}_n = \{ \widetilde{\psi}_{k,n} \}_{k=1}^N$ is a collection of bounded sequences in $\Hhalf(\RR^3)$. Hence, by passing to a subsequence, we find that $\widetilde{\psi}_{k,n} \rightharpoonup \widetilde{\psi}_{k,*}$, weakly in $\Hhalf(\RR^3)$, as well as $\widetilde{\psi}_{k,n}(x) \rightarrow \widetilde{\psi}_{k,*}(x)$ pointwise, for almost every $x \in \RR^3$, as $n \rightarrow \infty$. Correspondingly, we write $\bm{\widetilde{\Psi}}_* = \{ \widetilde{\psi}_{k,*} \}_{k=1}^N$, and the sequence of density matrices obeys $\rho_{\bm{\widetilde{\Psi}}_n}(x,y) \rightarrow \rho_{\bm{\widetilde{\Psi}}_*}(x,y)$ pointwise, for almost every $x$ and $y$ in $\RR^3$, as $n \rightarrow \infty$. Next, we claim that 
\begin{equation} \label{eq:strLp}
\rho_{\bm{\widetilde{\Psi}}_n}(x) \stackrel{n \rightarrow \infty}{\longrightarrow} \rho_{\bm{\widetilde{\Psi}}_*}(x), \quad \mbox{strongly in $L^p(\RR^3)$, for $1 < p < 3/2$,}
\end{equation} 
after possibly passing to a subsequence. To prove (\ref{eq:strLp}), we define the functions
\begin{equation}
\kappa_n(x) := \sqrt{\rho_{\bm{\widetilde{\Psi}}_n}(x)} = \sqrt{ \sum_{k=1}^N |\widetilde{\psi}_{k,n}(x)|^2} .
\end{equation} 
One can show (using \cite[Theorm 7.13]{Lieb+Loss2001}) that 
\begin{equation}
\| \kappa_n \|_{H^{1/2}}^2 \lesssim \sum_{k=1}^N \| \widetilde{\psi}_{k,n} \|_{H^{1/2}}^2 . 
\end{equation}
Thus, $\{ \kappa_n \}_{n \in \NN}$ forms a bounded sequence in $\Hhalf(\RR^3)$ and, by hypothesis of Theorem \ref{th:Chandra} and Remark 4) following Theorem \ref{th:Chandra}, we actually have that $\{ \kappa_n \}_{n \in \NN} \subset \Hrhalf(\RR^3)$, which denotes the subspace of spherically symmetric functions in $\Hhalf(\RR^3)$. Since the embedding 
\begin{equation}
\Hrhalf(\RR^3) \hookrightarrow L^p(\RR^3)
\end{equation} 
is compact if and only if $2 < p <3$ (see \cite{Sickel2000}), we deduce (after passing to a subsequence) that $\kappa_n \rightarrow \kapcr$, as $n \rightarrow \infty$, strongly in $L^p(\RR^3)$, for every $2 < p < 3$. This proves our claim (\ref{eq:strLp}).

Combining all the convergence properties shown above, we infer that
\begin{align}
\lim_{n \rightarrow \infty} \int_{\RR^3} \! \int_{\RR^3} \frac{ \rho_{\bm{\widetilde{\Psi}}_n}(x) \rho_{\bm{\widetilde{\Psi}}_n}(y) - |\rho_{\bm{\widetilde{\Psi}}_n}(x,y)|^2}{|x-y|} \, \dd x \, \dd y & \phantom{=}  \nonumber \\  =   \int_{\RR^3} \! \int_{\RR^3} \frac{ \rho_{\bm{\widetilde{\Psi}}_*}(x) \rho_{\bm{\widetilde{\Psi}}_*}(y) - |\rho_{\bm{\widetilde{\Psi}}_*}(x,y)|^2}{|x-y|} \, \dd x \, \dd y . \label{eq:strLP2}
\end{align}   
Here we have used (\ref{eq:strLp}) together with the Hardy-Littlewood-Sobolev inequality, as well as the dominated convergence theorem combined with the pointwise estimate $|\rho_{\bm{\widetilde{\Psi}}_n}(x,y)|^2 \leq \rho_{\bm{\widetilde{\Psi}}_n}(x) \rho_{\bm{\widetilde{\Psi}}_n}(y)$, which follows from the Cauchy-Schwarz inequality. 

We note, further, that 
\begin{equation} \label{eq:Enzzero}
0=\lim_{n \rightarrow \infty} \Enz(\bm{\widetilde{\Psi}}_n) \geq \Enz(\bm{\widetilde{\Psi}}_*),
\end{equation}
by using (\ref{eq:strLP2}) and the weak lower semicontinuity of the first term in (\ref{eq:Enz}). In addition, we obtain that $0 \leq \langle \widetilde{\psi}_{k,*} , \widetilde{\psi}_{l,*} \rangle \leq \delta_{kl}$ (in the sense of Hermitian $N \times N$-matrices), since $\widetilde{\psi}_{k,n} \rightharpoonup \widetilde{\psi}_{k,*}$ weakly in $L^2(\RR^3)$. Invoking Lemma 2 (see Appendix A) for $\bm{\widetilde{\Psi}}_* = \{ \widetilde{\psi}_{k,*} \}_{k=1}^N$ we find (similar to the proof of Theorem \ref{th:gwp}) that
\begin{equation}
\Enz(\bm{\widetilde{\Psi}}_*) \geq K \Big [ 1 - \frac{\kappa}{\kapcr}  \big ( \int_{\RR^3} \rho_{\bm{\widetilde{\Psi}}_*}(x) \, \dd x \big )^{2/3} \Big ]  \int_{\RR^3} \rho_{\bm{\widetilde{\Psi}}_*}(x)^{4/3} \, \dd x  ,  \label{eq:Daub}
\end{equation} 
where $K \geq 1.63$ is the constant from Lemma 2, and $\kapcr >0$ denotes the same universal constant as in Theorem \ref{th:gwp}. Moreover, we deduce from (\ref{eq:strLP2}) and (\ref{eq:hls}) that $\rho_{\bm{\widetilde{\Psi}}_*}(x) \not \equiv 0$ must hold. Thus, we see that inequalities (\ref{eq:Daub}) and (\ref{eq:Enzzero}) imply 
\begin{equation} \label{eq:minChandra}
\int_{\RR^3} \rho_{\bm{\widetilde{\Psi}}_*}(x) \, \dd x \geq \left ( \frac{\kapcr}{\kappa} \right )^{3/2} .
\end{equation} 

By using (\ref{eq:minChandra}), we find that assumption (\ref{eq:assChandra}) leads to a contradiction as follows. In view of (\ref{eq:strLp}), we deduce that $\rho_{\bm{\widetilde{\Psi}}_n}(x) \rightarrow \rho_{\bm{\widetilde{\Psi}}_*}(x)$ strongly in $L^1_{\mathrm{loc}}(\RR^3)$, as $n \rightarrow \infty$. For every $A > 0$, we thus obtain that
\begin{align}
 \int_{|x| < A} \rho_{\bm{\widetilde{\Psi}}_*}(x) \, \dd x & = \lim_{n \rightarrow \infty} \int_{|x| < A} \rho_{\bm{\widetilde{\Psi}}_n}(x) \, \dd x \nonumber \\
 & = \lim_{n \rightarrow \infty} \int_{|x| < \sigma(t_n)^{-1} A} \rho_{\TPsi(t_n)}(x) \, \dd x \nonumber \\
 & \leq \liminf_{n \rightarrow \infty} \int_{|x| < R}   \rho_{\TPsi(t_n)}(x) \, \dd x, \label{eq:Chandra33}
\end{align}
where we use that $\sigma(t_n)^{-1} \rightarrow 0$ as $n \rightarrow \infty$. Since estimate (\ref{eq:Chandra33}) holds for every $A > 0$, we deduce from (\ref{eq:minChandra}) that assumption (\ref{eq:assChandra}) leads to a contradiction. This completes the proof of Theorem \ref{th:Chandra}. \hfill $\blacksquare$


\begin{appendix}

\section{Lower bound for Kinetic Energy}

The following result is a slight extension of an estimate derived in \cite{Daubechies1983}. 

\begin{lemma} \label{lem:kin_energy}
Suppose $\TPhi = \{ \phi_k \}_{k=1}^N \subset \Hhalf(\RR^3)$ satisfies $0 \leq \langle \phi_k , \phi_l \rangle \leq \delta_{kl}$ in the sense of Hermitean $N \times N$-matrices. Then
\[
\sum_{k=1}^N \langle \phi_k, \sqrt{-\Delta} \, \phi_k \rangle \geq K \int_{\RR^3} \rho_{\TPhi}(x)^{4/3} \, \dd x,
\] 
where $\rho_{\TPhi}(x) = \sum_{k=1}^N |\phi_k(x)|^2$ and $K \geq 1.63$ is some constant.
\end{lemma}

\noindent
{\em Proof.} First we remark that both sides of the inequality to be shown are invariant under unitary transformations of the $\phi$'s, \ie, under the transformations $\phi_k \mapsto \sum_{l=1}^N A_{kl} \phi_l$, for an arbitrary unitary matrix $A \in U(N)$. Therefore we can assume without loss of generality that 
\begin{equation}
 \langle \phi_k, \phi_l \rangle = \lambda_k \delta_{kl}, \quad \mbox{with $0 < \lambda_k \leq 1$,}
\end{equation} 
where we have also discarded any possible zero vector, $\phi_k \equiv 0$, corresponding to $\lambda_k = 0$.

Next, we consider the $N \times N$-matrix, $H = (h_{kl})_{1 \leq k,l \leq N}$, with entries
\begin{equation}
h_{kl} = \langle \phi_k, \sqrt{-\Delta} \, \phi_l \rangle - c \langle \phi_k, U \phi_l \rangle .
\end{equation}
Here $U(x) := \rho_{\Phi}(x)^{1/3}$, and $c > 0$ is some constant to be chosen below. Since $H$ is Hermitean, there exists $B \in U(N)$ such that $\widetilde{H} = B^* H B$ has entries
\begin{equation}
\widetilde{h}_{kl} = \widetilde{\epsilon}_{k} \delta_{kl},
\end{equation}
with eigenvalues $\widetilde{\epsilon}_1 \leq \widetilde{\epsilon}_2 \leq \ldots \leq \widetilde{\epsilon}_N$. Let $\{ E_j\}_{j \geq 0}$ denote the set of negative eigenvalues of the relativistic Schr\"odinger operator $\sqrt{-\Delta} - cU$, acting on $L^2(\RR^3)$, and we consider the set of orthogonal vectors given by $\{ \widetilde{\phi}_k : \widetilde{\phi}_k = \sum_l B_{kl} \phi_l, \; \widetilde{\epsilon}_k < 0 \}$. Noting that $0 < \langle \widetilde{\phi}_k, \widetilde{\phi}_k \rangle \leq 1$ holds, we deduce that
\begin{align}
\sum_{k=1}^N h_{kk} & = \sum_{k=1}^N \widetilde{\epsilon}_{k}  \geq \sum_{{ k=1 \atop \widetilde{\epsilon}_k < 0}}^N \widetilde{\epsilon}_k \geq \sum_{{k=1 \atop \widetilde{\epsilon}_k < 0}}^N \big ( \frac{\widetilde{\epsilon}_k}{ \langle \widetilde{\phi}_k, \widetilde{\phi}_k \rangle } \big ) \nonumber \\ & \geq \sum_{j \geq 0} E_j \geq -L c^4 \int_{\RR^3} U(x)^4 \, \dd x. \label{eq:daub}
\end{align}
Here the first inequality in the second line follows from the min-max principle applied to the set of orthonormal vectors $\{ \widetilde{\phi}_k/ \| \widetilde{\phi}_k \|_{L^2} \}$ and the operator $\sqrt{-\Delta} - cU$. Moreover, the last inequality in (\ref{eq:daub}) is a standard estimate, where $L > 0$ is some constant; see \cite{Daubechies1983}. By choosing $c=2^{-2/3} L^{-1/3}$, we complete the proof of Lemma \ref{lem:kin_energy}, where the lower bound $K = 3/4 \cdot 2^{-2/3} L^{-1/3} \geq 1.63$ follows from known bounds for $L$. \hfill $\blacksquare$

\end{appendix}

\bibliographystyle{amsplain}

\begin{thebibliography}{10}

\bibitem{Cazenave2003}
Thierry Cazenave, \emph{Semilinear {S}chr\"odinger equations}, Courant Lecture
  Notes in Mathematics, vol.~10, New York University Courant Institute of
  Mathematical Sciences, New York, 2003. \MR{MR2002047 (2004j:35266)}

\bibitem{Chandrasekhar1931}
Subrahmanyan Chandrasekhar, \emph{The maximum mass of ideal white dwarfs},
  Astrophys. J. \textbf{74} (1931), 81--82.

\bibitem{Cho+Ozawa2006}
Yonggeun Cho and Tohoru Ozawa, \emph{On the semi-relativistic {Hartree} type
  equation},  (2006), Hokkaido Univ. Preprint Series in Math. \# 773.

\bibitem{Daubechies1983}
Ingrid Daubechies, \emph{An uncertainty principle for fermions with generalized
  kinetic energy}, Comm. Math. Phys. \textbf{90} (1983), no.~4, 511--520.
  \MR{MR719431 (85j:81008)}

\bibitem{EES2004}
Alexander Elgart, L{\'a}szl{\'o} Erd{\H{o}}s, Benjamin Schlein, and Horng-Tzer
  Yau, \emph{Nonlinear {H}artree equation as the mean field limit of weakly
  coupled fermions}, J. Math. Pures Appl. (9) \textbf{83} (2004), no.~10,
  1241--1273. \MR{MR2092307 (2005e:81273)}

\bibitem{Froehlich+Lenzmann2006}
J{\"u}rg Fr{\"o}hlich and Enno Lenzmann, \emph{Blow-up for nonlinear wave
  equations describing {Boson Stars}}, To appear in Comm. Pure Appl. Math.
  (2006), Preprint: arXiv:math-ph/0511003.

\bibitem{LenzmannDiss2006}
Enno Lenzmann, \emph{Nonlinear dispersive equations describing {Boson} stars},
  (2006), ETH Dissertation No.~16572.

\bibitem{Lenzmann2006}
\bysame, \emph{Well-posedness for semi-relativistic {Hartree} equations of
  critical type}, To appear in Mathematical Physics, Analysis, and Geometry
  (2006), Preprint: arXiv:math.AP/0505456.

\bibitem{Lieb+Loss2001}
Elliott~H. Lieb and Michael Loss, \emph{Analysis}, second ed., Graduate Studies
  in Mathematics, vol.~14, American Mathematical Society, Providence, RI, 2001.
  \MR{MR1817225 (2001i:00001)}

\bibitem{Lieb+Thirring1984}
Elliott~H. Lieb and Walter~E. Thirring, \emph{Gravitational collapse in quantum
  mechanics with relativistic kinetic energy}, Ann. Physics \textbf{155}
  (1984), no.~2, 494--512. \MR{MR753345 (86g:81037)}

\bibitem{Lieb+Yau1987}
Elliott~H. Lieb and Horng-Tzer Yau, \emph{The {C}handrasekhar theory of stellar
  collapse as the limit of quantum mechanics}, Comm. Math. Phys. \textbf{112}
  (1987), no.~1, 147--174. \MR{MR904142 (89b:82014)}

\bibitem{Merle+Tsutsumi1990}
Frank Merle and Yoshio Tsutsumi, \emph{{$L\sp 2$} concentration of blow-up
  solutions for the nonlinear {S}chr\"odinger equation with critical power
  nonlinearity}, J. Differential Equations \textbf{84} (1990), no.~2, 205--214.
  \MR{MR1047566 (91e:35194)}

\bibitem{Sickel2000}
Winfried Sickel and Leszek Skrzypczak, \emph{Radial subspaces of {B}esov and
  {L}izorkin-{T}riebel classes: extended {S}trauss lemma and compactness of
  embeddings}, J. Fourier Anal. Appl. \textbf{6} (2000), no.~6, 639--662.
  \MR{MR1790248 (2002h:46056)}

\bibitem{Straumann2004}
Norbert Straumann, \emph{General relativity}, Texts and Monographs in Physics,
  Springer-Verlag, Berlin, 2004, With applications to astrophysics.
  \MR{MR2069631 (2005m:83002)}

\bibitem{Weinstein1989}
Michael~I. Weinstein, \emph{The nonlinear {S}chr\"odinger
  equation---singularity formation, stability and dispersion}, The connection
  between infinite-dimensional and finite-dimensional dynamical systems
  (Boulder, CO, 1987), Contemp. Math., vol.~99, Amer. Math. Soc., Providence,
  RI, 1989, pp.~213--232. \MR{MR1034501 (90m:35188)}

\end{thebibliography}

\end{document}